\documentclass[article,twocolumn,showpacs,superscriptaddress,nofootinbib]{revtex4-2}

\usepackage[utf8]{inputenc}

\usepackage[colorinlistoftodos]{todonotes}
\usepackage{hyperref}
\usepackage{mathrsfs,amsmath}
\usepackage{cleveref}
\usepackage{subfiles}
\graphicspath{{figures/}{figuresSI/}{../figuresSI/}{../figures/}}
\usepackage{subfiles}
\usepackage{makeidx}
\usepackage{algorithm}% http://ctan.org/pkg/algorithms
\usepackage{algpseudocode}% http://ctan.org/pkg/algorithmicx

\algnewcommand{\LineComment}[1]{\State \(\triangleright\) #1}

\usepackage[normalem]{ulem}
\usepackage[super]{nth}
\usepackage{chngcntr}
\usepackage{subfiles} % Best loaded last in the preamble
\newcommand\longempty{}%
\usepackage{xr}
\newcommand\DoIfAndOnlyIfStandAlone{%
  \ifx\document\longempty
    \expandafter\@gobble
  \else
    \expandafter\@firstofone
  \fi
}%

\usepackage{tikz}
\usepackage{subfiles} % Best loaded last in the preamble
\usepackage{cleveref}
\usepackage{color}
\usepackage{booktabs}
\usepackage{mathtools}
\usepackage{cancel}
\graphicspath{ {./figures/} }
\usepackage{bbm}
\DeclareMathOperator*{\argmin}{argmin}   % Jan Hlavacek
\DeclarePairedDelimiterX{\infdivx}[2]{(}{)}{%
  #1\;\delimsize\|\;#2%
}
\newcommand{\infdiv}{\mathbb{D}\infdivx}

\usepackage{physics}
\usepackage{amssymb}
\usepackage{bm}% bold math
\usepackage{hyperref}% add hypertext capabilities
\newcommand{\be}{\begin{equation}}\newcommand{\ee}{\end{equation}}

\Crefname{equation}{Eq.}{Eqs.}
\Crefname{figure}{Fig.}{Figs.}
\Crefname{tabular}{Tab.}{Tabs.}

\crefname{equation}{Eq.}{Eqs.}
\crefname{figure}{Fig.}{Figs.}
\crefname{tabular}{Tab.}{Tabs.}

\usepackage[normalem]{ulem}

\begin{document}

\title{Variational dynamics of open quantum systems in phase space}
\author{Debbie Eeltink}
%\email{Corresponding authors: D. Eeltink: eeltink@mit.edu}
\affiliation{Laboratory of Theoretical Physics of Nanosystems (LTPN), Institute of Physics, \'{E}cole Polytechnique F\'{e}d\'{e}rale de Lausanne (EPFL), 1015 Lausanne, Switzerland}
\affiliation{Center for Quantum Science and Engineering, \\ \'{E}cole Polytechnique F\'{e}d\'{e}rale de Lausanne (EPFL), CH-1015 Lausanne, Switzerland}

\author{Filippo Vicentini}
\affiliation{Center for Quantum Science and Engineering, \\ \'{E}cole Polytechnique F\'{e}d\'{e}rale de Lausanne (EPFL), CH-1015 Lausanne, Switzerland}
\affiliation{CPHT, CNRS, \'{E}cole polytechnique, Institut Polytechnique de Paris, 91120 Palaiseau, France}
\affiliation{Coll\`ege de France, Universit\'e PSL, 11 place Marcelin Berthelot, 75005 Paris, France
}

\author{Vincenzo Savona}
\affiliation{Laboratory of Theoretical Physics of Nanosystems (LTPN), Institute of Physics, \'{E}cole Polytechnique F\'{e}d\'{e}rale de Lausanne (EPFL), 1015 Lausanne, Switzerland}
\affiliation{Center for Quantum Science and Engineering, \\ \'{E}cole Polytechnique F\'{e}d\'{e}rale de Lausanne (EPFL), CH-1015 Lausanne, Switzerland}

\begin{abstract}

We present a method to simulate the dynamics of large driven-dissipative many-body open quantum systems using a variational encoding of the Wigner or Husimi-Q quasi-probability distributions.
The method relies on Monte-Carlo sampling to maintain a polynomial computational complexity while allowing for several quantities to be estimated efficiently.
As a first application, we present a proof of principle investigation into the physics of the driven-dissipative Bose-Hubbard model with weak nonlinearity, providing evidence for the high efficiency of the phase space variational approach.

\end{abstract}

\maketitle

\section*{Introduction}
The development of larger quantum devices for quantum technology applications is increasingly defying the efficiency of theoretical models and simulation methods used to design and validate such devices \cite{bruzewicztrapped19,ciracquantum20,kjaergaardsuperconducting20,browaeysmany-body20}. 
%\sout{One of the most challenging aspects features of these systems is that they} 
These devices are open, namely subject to the influence of the surrounding environment, both as a nuisance and as a way of control, making it possible to develop quantum correlations and mixedness through the action of the environment, which can steer the dynamics of the system \cite{rotterReviewProgressPhysics2015,kochControllingOpenQuantum2016,LeghtasScience15, deNeeve2022NatPhysDissipativePumpingCorrection}. %\sout{calls for efficient methods for the simulation of open quantum systems \cite{weimerSimulationMethodsOpen2021}.} 

The mixed nature of the state describing an open quantum system arises by the tracing out of the environment degrees of freedom. 
This state is then usually encoded as a (positive semidefinite) density matrix.
The dynamics can then be described by the Lindblad master equation (LME, \cref{eq:LME}) under the assumption of a memoryless (i.e., Markovian) bath weakly interacting with the system such that its state remains unperturbed (Born approximation) \cite{gardiner_zoller,breuer_theory_2007}. 
Three mainstream representations of such dynamics exist,  all retrieving the same expectation values of any operator: the integration of the LME, the unraveling of the density matrix through stochastic wave functions called quantum trajectories, and the integration of the differential equations describing quasi-probability functions in phase space.

First, solving the LME for a quantum system, defined in a $N$-dimensional Hilbert space, requires solving $N^2$ coupled linear ordinary differential equations (ODEs). 
Second, the unraveling of the master equation onto a statistical ensemble of stochastic quantum trajectories \cite{Molmer1993WFMC,breuerTheoryOpenQuantum2007,gardiner_zoller}, requires solving $N$ coupled linear stochastic differential equations (SDEs) multiple times until statistical convergence is established.
Both methods suffer the curse of dimensionality, as the number of differential equations to be solved scales exponentially with the number of modes, and calls for efficient numerical methods and approximations \cite{weimerSimulationMethodsOpen2021}.
These include mean-field \cite{jincluster16,Verstraelen2018GaussianTraj,verstraelen2023quantum} and linked-cluster expansions \cite{biellalinked18}, and corner-space renormalization \cite{Finazzi2015PRLCorner,donatellacontinuous-time21}.

The third possibility is the description of an open quantum system in terms of a quasi-probability distribution (QPD) in phase space \cite{carmichaelStatisticalMethodsQuantum1999,gardiner_zoller,wallsQuantumOptics2008,polkovnikovPhaseSpaceRepresentation2010,rundleOverviewPhaseSpace2021}. 
Then, a QPD is governed by a partial differential equation (PDE) whose dimensionality is set by the number of modes composing the system.
While QPDs are useful analytical tools, numerically integrating such partial differential equations with no further approximations is usually a much harder task than the original problem where fewer algorithms are available and most scale poorly as dimensionality increases. 
While some progress have been made in this direction \cite{veronezPhaseSpaceFlow2013,holmesHusimiDynamicsGenerated2023,roda-llordesNumericalSimulationLargeScale2023a}, the presence of higher-order terms in the differential equation, and in the  highly-singular nature of the  solutions adds considerable challenges \cite{drummondGeneralisedPrepresentationsQuantum1980,wallsQuantumOptics2008,gilchristPositiveRepresentationApplication1997}.
Only in specific cases or approximations, where governing PDE reduces to true a Fokker-Planck equation, have phase space methods been extensively investigated and adopted to simulate a variety of (multi-mode) models of open and closed quantum systems \cite{gilchristPositiveRepresentationApplication1997,deuarFullyQuantumScalable2021b,kiesewetterScalableQuantumSimulation2014,Vogel1989,sinatraTruncatedWignerMethod2002,deuarGaugeRepresentationsQuantumdynamical2002,plimakQuantumfieldtheoreticalApproachPhasespace2003a,CarusottoRMP13,Foss-FeigPRA17,VicentiniPRA18,HuberPRA22}.
Indeed, such approximations are often limited to weakly-interacting, almost semi-classical systems.

Numerical methods that reduce the computational complexity of the problem and that can deal with strong interactions and non-classical states focused mainly on a discrete description of the LME, for example leveraging variational ansätze \cite{Weimer2015PRLVariationalprincipleopen,OverbeckPRA17}, including
tensor- \cite{Cui2015PRLMPSDiss,MascarenhasPRA2015MPS,wernerPositiveTensorNetwork2016,orusPracticalIntroductionTensor2014,weimerSimulationMethodsOpen2021} or neural-network representations \cite{nagyVariationalQuantumMonte2019,Hartmann2019PRLDissipative,vicentiniVariationalNeuralNetworkAnsatz2019,Yoshioka2019PRB} of the quantum state. 
Recent works have refined the latter method, either by ensuring that the ansatz is always physical for arbitrary network depths \cite{Vicentini22arxivPosDef} or by leveraging more efficient, nonphysical parametrizations \cite{rehTimeDependentVariationalPrinciple2021b,Luo2022}.

However,  generalizing such  variational methods  to the LME for bosonic systems in a truncated Fock space has proven challenging so far \cite{Saito2017BoseHubbard}.
To this day, an efficient and reliable approach to encoding the wave-function or density-matrix of an bosonic open system -- or more generally of a system with a very large local Hilbert space dimension -- on a neural network is not known.  One-hot encodings of the local degree of freedom to find the ground state have been tested \cite{Saito2018Bosons,even2022Bose2Leg}, but their computational cost is high \cite{Pei2021Spin1Encoding}.
For closed systems,  an efficient representation of a bosonic system was found by using a first-quantized description of the wave-function in the position basis \cite{Saito2018FirstQuantised,Pescia2022Bosons}, but the conservation of particle number cannot be directly translated to the dynamics of open systems.

Following this intuition, the phase space and its natural position-momentum variables emerges as a suitable platform to implement the variational principle for bosons or continuous variable systems. 
Indeed, the possibility of representing a true Fokker-Planck equation with variational ansätze has recently emerged \cite{rehVariationalMonteCarlo2022a}. Very recently, the same idea has been applied to QPDs of coupled linear modes \cite{duganQFlowGenerativeModeling2023}.

Here, we apply the time-dependent variational principle (tVMC) to parameterized phase space functions, and show that we can accurately simulate time evolution of driven-dissipative systems in the limit of small but nonvanishing nonlinearity, up to arbitrary occupation number. We apply the tVMC method using both a neural-network and a complex-Gaussian variational ansatz. We discuss the limitations of the adopted variational ansatz and of the size of the time step in the tVMC context. As the output of the model is the full QPD, quantum correlations among modes are efficiently described. Moreover, sampling the variational QPD allows for efficient calculation of expectation values of quantities such as occupation number, Wigner negativity and entropy. We show several proof-of-concept results involving Wigner negativity, non-Gaussian initial conditions, forcing, dissipation, and explore the limitations imposed by nonlinearity.

%============= RESULTS AND DISCUSSION =========
\section*{Results and discussion}

\begin{figure}
\includegraphics[width=8cm]{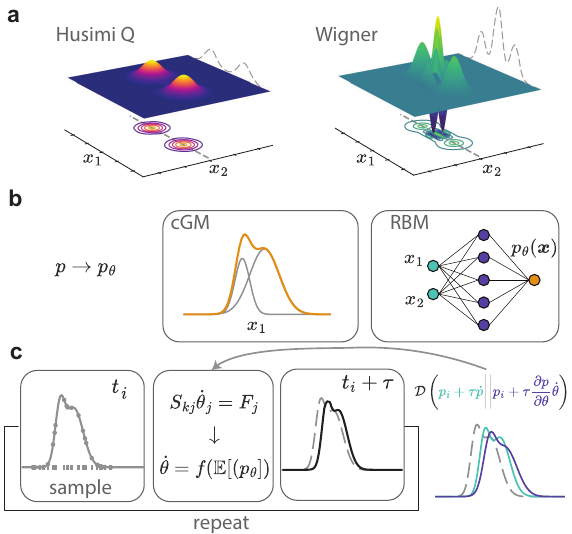}% Here is how to import EPS art
\caption{Sketch of the method discussed in this manuscript. \textbf{(a)} Plots of the Husimi-Q (left) and Wigner (right) QPDs for a cat state, as an illustration. \textbf{(b)}  The QPD $p$ in \cref{eq:FP} is parameterized by either a RBM or cGM ansatz. \textbf{(c)}  Illustrative flow chart of the tVMC method. The parameterized function $p_\theta$ is sampled. The optimal step in parameter space is given by the set of ODEs in which the $S$ matrix and $F$ vector can be efficiently calculated based on expectation values over the given samples.}\label{fig:method}
\end{figure}

\subsubsection*{Phase space representation}
We consider a quantum system consisting of M bosonic modes dissipatively coupled to a Markovian environment.
The dynamics is governed by the Lindblad master equation (LME) for the density matrix $\rho$ \cite{breuer_theory_2007,gardiner_zoller}.
Setting $\hbar=1$,
\begin{equation}\label{eq:LME}
    \dot{\rho}(t) = \mathcal{L} \rho(t) = -i [H,\rho]+ \sum_{k=1}^K\mathcal{D}[C_k]\rho(t)\, , 
\end{equation}
where $H$ is the system Hamiltonian and $\mathcal{L}$ denotes the Liouvillian superoperator which generates the non-unitary dynamics.
The dissipator $\mathcal{D}[C_k]$ is defined in terms of the jump operators $C_k$ as  $\mathcal{D}[C_k]\rho =\frac{1}{2}\left[2C_k \rho C_k^\dag - \rho C_k^{\dag}C_k-C_k^{\dag}C_k\rho\right]$.
%\begin{equation}\label{eq:dissipOperator}
%   \mathcal{D}[C_k]\rho =\frac{1}{2}\left[2C_k \rho C_k^\dag - \rho C_k^{\dag}C_k-C_k^{\dag}C_k\rho\right]\,. 
%\end{equation}
Phase space methods exactly map the state of the system on one of several possible quasi-probability distributions (QPDs) $p(\bm{x} ,t)$, and the dynamics equation~\eqref{eq:LME} onto a PDE that determines the evolution of the QPD.
The particular structure of the PDE is determined by the choice of QPD.
The most relevant choices are the Wigner, Glauber P, and Husimi-Q distributions, for which the mapping are reviewed in the SI.

Many-body interactions and dissipation can lead to a high-order PDE describing the function $p(\bm{x} ,t)$. 
Most applications of phase space methods truncate terms higher than second order, thereby yielding a 2M dimensional quasi Fokker-Planck (FP) equation characterized by a drift vector $\bm{\nu}(\bm{x},t) $ and diffusion tensor  $\mathbf{D}(\bm{x} ,t) $\cite{drummondQuantumTheoryNonlinear2014,gardinerQuantumWorldUltraCold2014}:
\begin{align}
\begin{split} \label{eq:FP}
  \frac{\partial p(\bm{x} ,t)}{\partial t}=&  -\sum_{i=1}^{2M} \frac{\partial}{\partial x_i} \left[ \nu_i(\bm{x} ,t) p(\bm{x} ,t) \right]\\
 + \sum_{i=1}^{2M} \sum_{j=1}^{2M}& \frac{\partial^2}{\partial x_i \, \partial x_j} \left[ D_{ij}(\bm{x} ,t) p(\bm{x} ,t) \right] 
\end{split}
\end{align}
 Here, $\bm{x} = (\Re{\alpha_1}, \Im{\alpha_1}...\Re{\alpha_M}, \Im{\alpha_M})$ for $M$ modes.
Since it is not possible to numerically integrate \cref{eq:FP} with standard solution techniques such as finite element or finite difference methods beyond three dimensions \cite{pichlerNumericalSolutionFokker2013,dobsonEfficientDatadrivenSolver2019,zhaiDeepLearningMethod2020},  stochastic differential equation (SDE) trajectory methods are usually employed \cite{carmichaelStatisticalMethodsQuantum1999,gardiner_zoller,wallsQuantumOptics2008,polkovnikovPhaseSpaceRepresentation2010,gilchristPositiveRepresentationApplication1997,deuarFullyQuantumScalable2021b,kiesewetterScalableQuantumSimulation2014,Vogel1989,sinatraTruncatedWignerMethod2002,deuarGaugeRepresentationsQuantumdynamical2002}. 
However, to do so, the PDE must be a true FP equation, meaning that $p(\bm{x},t)$ should be non-negative, and the diffusion matrix should be positive-semi-definite. 
These two requirements respectively rule out the negative Wigner function and the Husimi-Q function.
In addition, when long-range quantum correlations dominate over local dissipation, the number of trajectories to sample the QPD may grow exponentially and the SDEs become unstable \cite{deuarFullyQuantumScalable2021b}. 

\subsubsection*{Variational ansatz and tVMC}
Instead, in the following examples, we  use a variational approach to integrate \cref{eq:FP}, where the drift and diffusion operators depend on the parameters of the Hamiltonian, the dissipator, and the chosen QPD (Sec. Methods details the procedure to construct \cref{eq:FP}, and gives an example of a specific form in  \cref{eq:CatDecohQ_PDEmain}).
We follow a scheme similar to variational neural-network quantum states, which have been successfully used to compress the wave-function of quantum systems with discrete \cite{Carleo2017,Chen23MachinePrecision} or continuous degrees of freedom \cite{Saito2018FirstQuantised,Pescia2022Bosons,Lovato2022HiddenNucleon} onto a polynomially-large set of variational parameters $\bm{\theta}$.

In particular, we efficiently represent the logarithm of the Husimi-Q ($p(\bm{x} ,t) = Q(\bm{x} ,t)$) or Wigner ($p(\bm{x} ,t)=W(\bm{x} ,t)$) QPD with a variational function $\log p(\bm{x}; \bm{\theta}(t))$ which may in general not be normalized (\cref{fig:method}(a)).
The time-dependence is encoded in the parameters $\bm{\theta}(t)$, which are in a 1-to-1 correspondence with quantum states.
The details of the variational function are given in Sec. Methods.
We explore two ansätze: a complex Gaussian mixture (cGM) and a restricted Boltzmann machine (RBM) (\cref{fig:method}(b)). 
We approximately map the PDE onto a system of ODEs for $\bm\theta(t)$ by means of two variational principles for the $Q$ and $W$ QPDs \cite{Yuan2019, rehVariationalMonteCarlo2022a}.  
In both cases, we obtain the differential equation
\begin{equation}
    \label{eq:tdvp-equation-def}
    \frac{d\bm\theta(t)}{dt} = S_{\bm\theta(t)}^{-1}\bm{F}_{\bm\theta(t)},
\end{equation}
where the entries of the vector of so-called variational forces $\bm{F}_{\bm\theta(t)}$ and the quantum geometric tensor $S_{\bm\theta(t)}$ are defined as
\begin{align}
    \label{eq:tdvp-forces-definition}
    \left[\bm{F}_{\bm\theta(t)}\right]_{i} &= \int p(\bm{x})^r\frac{d \log p(\bm{x})}{dt}\frac{\partial \log p(\bm{x})}{\partial\theta_i} d\bm{x} \\
    \label{eq:tdvp-qgt-definition}
    \left[S_{\bm\theta(t)}\right]_{i,j} &= \int p(\bm{x})^r\frac{\partial \log p(\bm{x})}{\partial\theta_i}\frac{\partial \log p(\bm{x})}{\partial\theta_j} d\bm{x}.
\end{align}
(We drop the dependency of $p$ from the parameters $\theta(t)$ to lighten the notation.) 

To derive this equation, we minimize the distance $\mathbb{D}$ between the time evolution of the QPD and the parameter evolution $\dot{\theta}$,
\begin{equation}\label{eq:TDVP}
 \argmin_{\dot{\theta}}  \infdiv*{p(\bm{x})+\frac{\partial p(\bm{x})}{\partial t}dt }{p(\bm{x})+ \sum_{i} \frac{\partial p(\bm{x})}{\partial \theta}\frac{\partial \theta_{i}}{\partial t} dt}.
\end{equation}
It is essential to chose a distance $\mathbb{D}$ that leads to quantities that can be (i) estimated by sampling an expectation value, and (ii) have finite variance. 
In particular, the Kullback–Leibler or Hellinger distance is a valid choice for the Husimi-Q \cite{rehVariationalMonteCarlo2022a} as it is positive and we can directly sample it ($r=1$).
For the Wigner QPD, instead, we select the $L^2$ norm because it can be efficiently estimated by sampling $W^2(\bm{x})$ ($r=2$) \footnote{This is strictly equivalent to re-deriving the McLachlan variational principle \cite{Yuan2019} for the QPD}.

We then replace the integrals in \cref{eq:tdvp-forces-definition,eq:tdvp-qgt-definition} with unbiased stochastic estimates that require sampling the QPD $p(\bm{x}; \bm{\theta}(t))$ or some power of it \cite{Carleo17tVMC,Schmitt2020PRLNoiseReg,rehTimeDependentVariationalPrinciple2021b} (\cref{fig:method}(c)). 
We remark that this approach is conceptually different from what is known as Physics-Informed Neural Networks (PINNs), where a variational ansatz $\tilde{p}(\bm{\theta}; \bm{x} ,t)$ is taken to represent the state at all points in space and time and the continuity equation is enforced on a randomly sampled grid \cite{Chen18NeuripsNODE,Cai2021PINN1,Cai2021PINN2}.

\subsubsection*{Single-mode benchmark}
\begin{figure*}
\includegraphics[width=16.5cm]{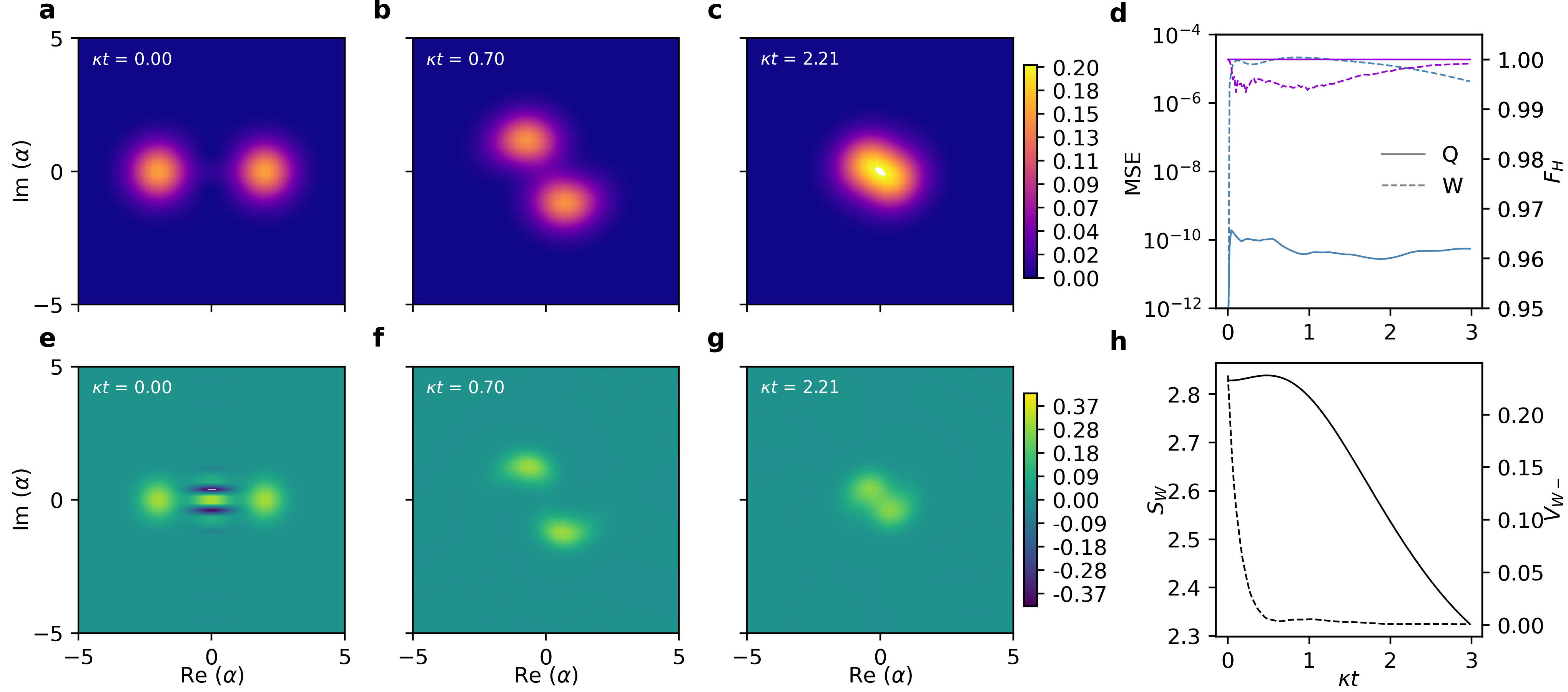}% Here is how to import EPS art
\caption{\label{fig:catdecoh} Simulation of nonlinear decoherence of cat state using a cGM ansatz with 16 components for the Husimi-Q and Wigner functions respectively, using parameters $\Delta/\kappa = 2$, $U/\kappa=0.1$ in \cref{eq:CatHam}. Time evolution snapshots of the Husimi-Q function  \textbf{a-c}  and the Wigner function \textbf{e-g} . \textbf{d}  MSE (green) for the Wigner (dashed) and Q (solid) function. Fidelity $F_\text{H}$ (purple) for Wigner (dashed) and Husimi-Q (solid).  \textbf{h}  Wehrl entropy $S_W$ (dashed) for the Husimi-Q function, and relative negative Wigner volume $V_{\text{W-}}$ (solid).}
\end{figure*}
We first consider a system of one anharmonic oscillator governed by the Hamiltonian 
\begin{equation}\label{eq:CatHam}
    H = \Delta a^\dag a  - U a^\dag a^\dag a a  .
\end{equation}
and dissipator $\mathcal{D}[\sqrt{\kappa}a]$. Here $a$ is the bosonic annihilation operator, obeying $[a,a^\dagger]=1$, $\Delta$ and $U$ are respectively the bare energy and anharmonic strength of the oscillator, and $\kappa$ is the dissipation rate.
In what follows, we take $\Delta/\kappa = 2$ and $U/\kappa=0.1$ and assume an initial Schr\"{o}dinger cat state $\ket{\mathcal{C}_\pm} = \ket{\alpha} \pm \ket{-\alpha}$.
We simulate the dynamics using a cGM ansatz with $R=16$ Gaussian components.
Snapshots of the Wigner and Husimi-Q functions are shown in \Cref{fig:catdecoh} (a-c) and (e-g) respectively.
We numerically show that our variational approach can describe the departure from the cat state, originating from the anharmonic effects.
The accuracy of the simulation is assessed by comparing against a numerically exact solution of the Lindblad master equation~\cite{johanssonQuTiPPythonFramework2013}.
\Cref{fig:catdecoh}(d) and (h) respectively show the mean squared error and a fidelity defined as $F_\text{H} = 1-\mathbb{D}_\text{H,abs} = \sqrt{|p_\textsc{ME}(\bm{x})||p_\theta(\bm{x})|} $ in terms of the Hellinger distance $\mathbb{D}_\text{H,abs}$.
Based on these metrics, the variational Husimi-Q function represents the state of the system almost exactly, while the truncated Wigner approximation departs slightly from the exact solution due to the anharmonic terms that are approximated by the truncation.

To prove the predictive value of our method, we also show that we can cheaply estimate quantities such as the Wehrl entropy by using Monte-Carlo integration, which are otherwise computationally intractable for trajectory methods.
The Wehrl entropy \cite{wehrlRelationClassicalQuantummechanical1979} for the Husimi-Q function is defined as
\begin{equation}
    S_W = -\int Q (\bm{x})\ln[Q(\bm{x})] \ d\bm{x} = -\mathbb{E}_{\bm{x}\sim Q(\bm{x})}[\ln Q  (\bm{x})],
\end{equation}
where the last identity shows how it can be naturally estimated by sampling the QPD. 
The quantity $S_W$, plotted in \cref{fig:catdecoh}(h), decreases along the dynamics, as expected for a purely dissipative system.

For the Wigner function, the negative volume is defined as $V_\text{W-} = \int W(\bm{x}) C(\bm{x}) d\bm{x} = \mathbb{E}_{\bm{x}\sim |W(\bm{x})|}[C(\bm{x})]$, where $C(\bm{x})= 1$ if $W(\bm{x}) < 0$ \cite{pizzimentiComplexvaluedNonGaussianityMeasure2023}.
The quantity $V_\text{W-}$, shown in \cref{fig:catdecoh}(h), vanishes as the cat state decays to the vacuum. 

\begin{figure}[ht]
\includegraphics[width=8cm]{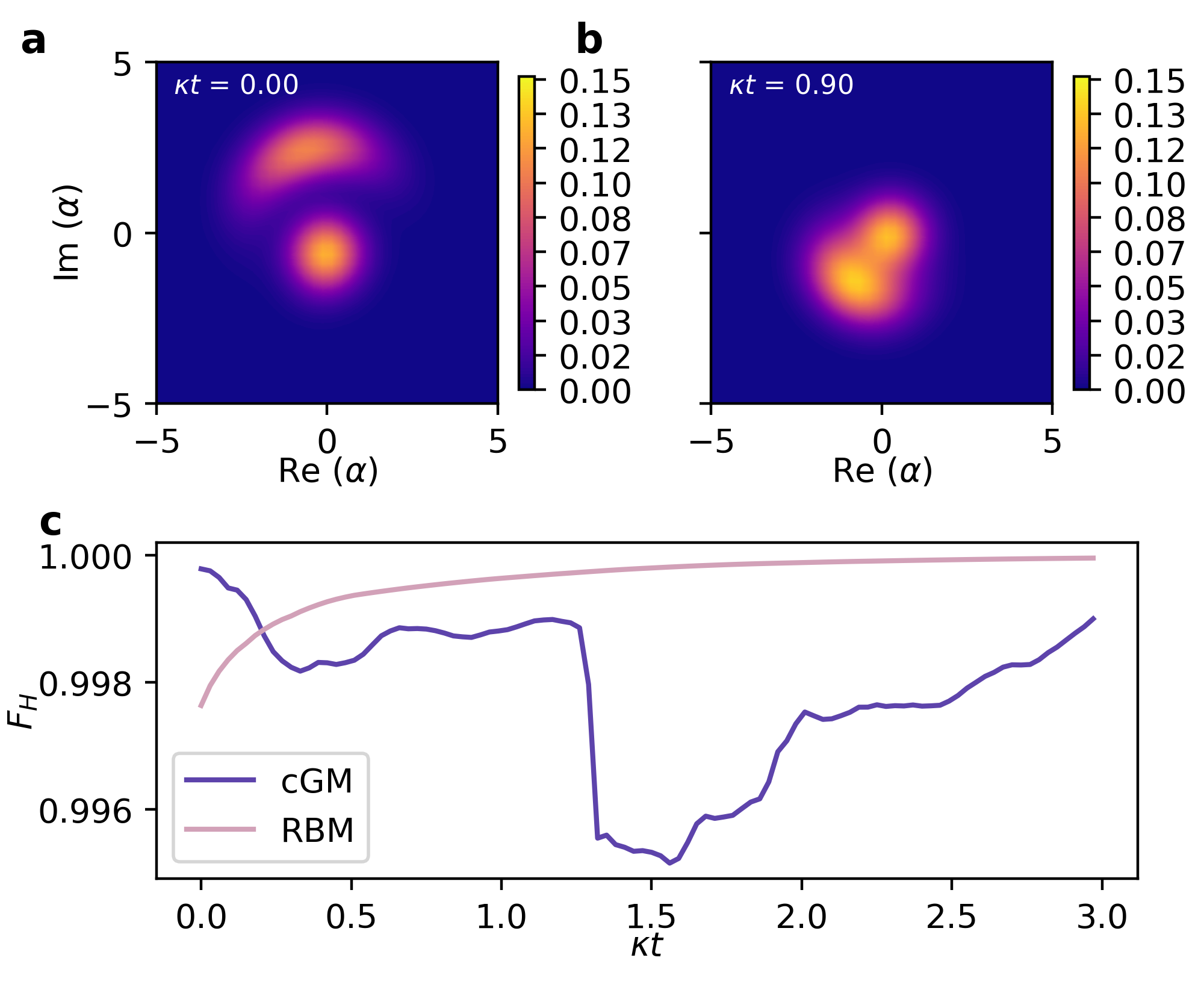}
\caption{Simulation starting from a density matrix corresponding to a bistable state, for cGM (8 components) and RBM ($\beta=30$) ansatz. \textbf{a,b}  Snapshots of the evolution of the Q function. Evolution parameters: $\Delta/\kappa = 2$, $F/\kappa=0.3$ $U/\kappa=0.07$. \textbf{c} evolution of $F_H$ for the cGM (pink) and RBM (purple) ansatz} \label{fig:bistabIC} 
%[WITH THE CGM I CAN PUSH TO: $\Delta = \pi$, $F=4$ and one photon dissipation rate $\kappa=\frac{1}{2}\pi$., WITH THE RBM I CANNOT]. 
\end{figure}

We now consider a driven-dissipative system with Hamiltonian
\begin{equation}\label{eq:HamBistable}
    H = \Delta a^\dag a  - U a^\dag a^\dag a a +F (a^\dag+ a), 
\end{equation}
and the same dissipator as before. This description of a Kerr resonator is a hallmark example in quantum optics \cite{Vogel1989,gardiner_zoller}.  Here, $F$ is the (real) driving field amplitude, and the Hamiltonian is expressed in the rotating frame of the drive, so that $\Delta$ represents the frequency detuning between the mode and the driving field. 
For this case, we set the initial state to be an arbitrary mixed state.
Specifically, we choose a state in the bistability region of parameters of the Kerr-resonator \cite{Vogel1989,drummondGeneralisedPrepresentationsQuantum1980}. The initial state is chosen so as to differ from the steady state for the given system parameters.
The results are displayed in \cref{fig:bistabIC}.
Both ansätze accurately describe the non-Gaussian pattern in the QPD arising along the dynamics, as seen in the time snapshots of \cref{fig:bistabIC}(a,b).
In \cref{fig:bistabIC}(c), the fidelity $F_\mathrm{H}$ computed along the dynamics is displayed, showing that both the RBM and the cGM ansätze efficiently represent the QPD dynamics. 

% \subsection{Coupled modes}
  \begin{figure*}[ht]
\includegraphics[width=18cm]{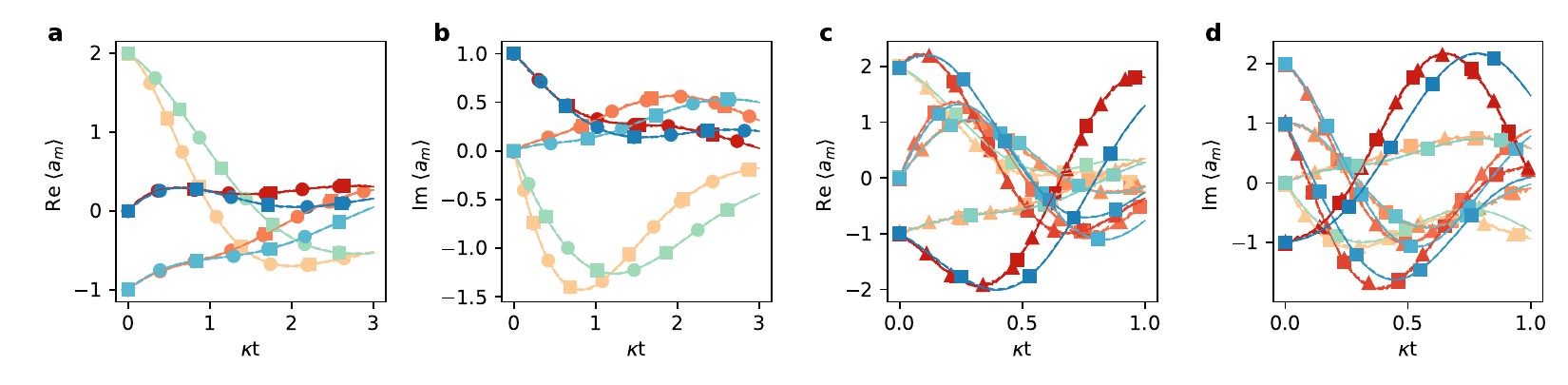}
\caption{Dynamics of several coupled modes. Real \textbf{a,c}  and imaginary \textbf{b,d}  part of $\langle a_m \rangle$.
% $\langle a_m \rangle = \alpha_m$
Blue shaded lines: $U/\kappa=0$, red shaded lines: $U/\kappa=0.1$. \textbf{a,b}  $M=3$, with parameters $\Delta/\kappa =  1$,  $J/\kappa = 0.3$, and $F/\kappa=0.3$ on the first mode only. Initial condition: $\psi_0 = \ket{\alpha_1}\otimes\ket{\alpha_2}\otimes\ket{\alpha_3}$ with values $\alpha_{1,2,3}=[2,1,i]$. RBM ansatz with $\beta=20$ (squares), 6-component cGM (circles), and LME (solid). \textbf{c,d}  $M=6$ with parameters $\Delta/\kappa = 4$, $J/\kappa =1$, and $F=0$. Initial condition: tensor product of coherent states, with displacement values $\alpha_{1..6}=[2,1,i,2i,2+i,-1-i]$.  RBM ansatz with  $\beta=5$ at solver step $\kappa dt = 10^{-4}$ (squares) and $\kappa dt = 2\times 10^{-5}$ (triangles). Second-order truncated cumulant expansion (solid). \label{fig:coupledModesObs} }
\end{figure*}
\subsubsection*{Several coupled modes}
Simulating the dynamics of several coupled modes provides evidence for the efficiency of the method. We first perform simulations on $M=3$ coupled modes -- a system that can still be reasonably simulated by direct integration of the Lindblad master equation on a truncated Fock space.
The system is characterized by the Hamiltonian 
\begin{align}
\begin{split}\label{eq:ham:3modeHop}
     H = &\sum_i^M (\Delta a_i^\dag a_i   - U a_i^\dag a_i^\dag a_i a_i)    +  F (a_1^\dag+ a_1)  \\
    &+ J\left[ \sum_{j=1}^{M-1}(a_{j+1}^\dag a_j   + a_j^\dag a_{j+1} )   +  a_{1}^\dag a_M   + a_M^\dag a_{1}  \right]
\end{split}
\end{align}
The modes are coupled through nearest neighbor hopping with periodic boundary conditions. Here, $J$ is the coupling constant, $U$ is the strength of the Kerr nonlinearity, $\Delta$ is the frequency detuning of the driving field, and $F$ is the driving field strength.
All modes are coupled to the environment through the dissipator $\mathcal{D}[\sqrt{\kappa}a]$. We assume that only the first mode is driven with $F/\kappa=0.3$.
The system is initialized to a coherent state in each mode, $\psi_0 = \ket{\alpha_1}\otimes\ket{\alpha_2}\otimes\ket{\alpha_3}$ with values $\alpha_1=2$, $\alpha_2=1$ and $\alpha_3=i$.
While our method gives the evolution of the full continuous 6 dimensional Husimi-Q function, expectation values of physical observables can again be efficiently obtained through sampling.

\Cref{fig:coupledModesObs} shows the evolution the real and imaginary parts of the field expectation values $\langle a_m\rangle$, with $U/\kappa = 0.1$. The RBM and the cGM ansatz closely follow the direct integration of the LME. The same quantities for $U=0$ are also displayed in \Cref{fig:coupledModesObs}, showing the strong influence of the nonlinearity on the evolution. 

We now consider a higher number of coupled modes, for which the direct solution of the LME is no longer feasible with reasonable computational means. To assess the accuracy of the phase space tVMC method, we compare the results to those obtained via truncated cumulant expansion \cite{kuboGeneralizedCumulantExpansion1962,plankensteinerQuantumCumulantsJlJulia2022}.
\Cref{fig:coupledModesObs}(c,d) shows the dynamics of the field expectation values for $M=6$ coupled modes governed by the Hamiltonian \cref{eq:ham:3modeHop}. The initial state is again set to a tensor product of coherent states, with values $\alpha_{1..6}=[2,1,i,2i,2+i,-1-i]$. For $U/\kappa =0$, the simulation using an RBM ansatz closely follows the exact result of the first order cumulant expansion or Gross-Pitavskii equation. For a finite nonlinearity $U/\kappa =0.1$, the accuracy of the variational phase space method is confirmed by solving for two different time steps $dt = 10^{-4}$ and $dt = 2\times10^{-5}$, and comparing to the second order cumulant expansion. 

%% == Figure Performance ===
\begin{figure}
\begin{center} 
\includegraphics[width=7.6cm]{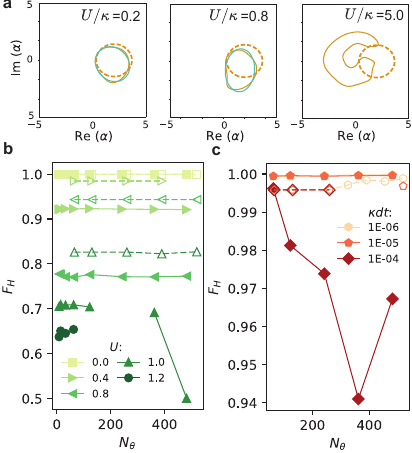}% Here is how to import EPS art
\caption{\label{fig:perf} Influence of nonlinearity on the solver performance. \textbf{a} Contour-plots of the Husimi-Q QPD corresponding to $Q(\alpha,\alpha^*)=0.05$, computed at $\kappa t=0.1$ and $U/\kappa = 0.2, 0.8$ and $5.0$. Dashed line: initial condition. Solid yellow: LME solution, green: cGM ansatz (for $U/\kappa = 5.0$ the numerical integration becomes unstable). \textbf{b,c}  $F_H$ as a function of number of parameters $N_\theta$ for the RBM ansatz (filled, solid) and cGM ansatz (open, dashed), for \textbf{b} $U/\kappa=0,\,0.4,\,0.8,\,1.0,\,1.2$. \textbf{c}  solver step sizes $\kappa dt =10^{-4},\,10^{-5}$, and $10^{-6}$ for $U/\kappa = 0.8$}
\end{center}
\end{figure}

\subsubsection*{Influence of nonlinearity.} 
In order to assess the influence of nonlinear terms on the predictive power of the method, we study the simplest case of a single mode undergoing the competition of nonlinearity and dissipation (see SI for more details). We set an initial coherent state $\ket{\alpha = 2 +0i}$. As the nonlinear term $U$ induces a departure from a Gaussian pattern in phase space, we expect the cGM to become rapidly less expressive as $U$ is increased. The RBM ansatz is more flexible, but we still expect difficulties due to the Gaussian tail-behavior that we have assumed. 
\Cref{fig:perf}(a) shows the departure from a Gaussian shape as $U$ is increased, at a fixed propagation time $\kappa\Delta t = 0.1$. 

A higher value of $U/\kappa$ results in a lower fidelity, and eventually leads to a numerical instability of the solver, for both the RBM and the cGM.  Increasing the number of parameters does not improve the performance. However, we observe that decreasing the time step drastically improves the performance. 

%Indeed, as \cref{eq:TDVP} is a first-order equality, when time-step is too large, the variational dynamics starts to detach from the exact solution. Differently put, in the parameter-space, the ansatz is stepping out of the manifold that is compatible with the PDE. Another reason this detachment occurs is when the ansatz cannot represent the exact solution.

Indeed, in presence of nonlinearity, the PDE \cref{eq:FP} becomes stiff and the first-order integration scheme used to treat integrate \cref{eq:tdvp-equation-def} requires an increasingly small time-step to yield accurate results.
We expect that higher-order, adaptive time-stepping scheme would mitigate the problems, yet, a straightforward application such as that discussed in Ref.~\cite{Schmitt2020PRLNoiseReg} does not help.
Alternatively, recently proposed implicit integration schemes solving a non-convex optimisation problem at each time-step, which allow for non-smooth evolutions along the variational manifold, might be more stable \cite{sinibaldiUnbiasingTimedependentVariational2023,donatella2022dynamics,gutierrez2022real}.

\section*{Conclusions.}
We have demonstrated that the dynamics of driven-dissipative open quantum systems made of several coupled bosonic modes can be efficiently integrated in phase space, by introducing a parameterized ansatz for the  Wigner or Husimi-Q QPDs, and adopting the tVMC method for their evolution. In this way, and by choosing an appropriately expressive variational ansatz, the large dimensionality of the many-body problem can be drastically reduced. Specifically, we turn a high dimensional PDE into a set of ODE's, at the cost of having to sample the the parameterized function. We provided empirical evidence that the variational phase space approach is particularly efficient in cases where the bosonic modes are characterized by a weak nonlinearity. We simulated the dynamics of several models with various initial conditions, giving a proof of principle of the efficiency of the variational phase space approach. An advantage of variationally representing the QPD in phase space is that some relevant observables, such as various entropy definitions, are easily accessible. Numerical simulations suggest that the main limitations of the present method are the representative power of the ansatz and the size of the time step, similarly to most tVMC approaches. In particular, these factors set an upper bound to the strength of nonlinearity that can be accommodated. The exploration of different, more complex ansätze, as well as the inclusion of an adaptive time step approach \cite{sinibaldiUnbiasingTimedependentVariational2023} are therefore a prospect for future research. In addition, the question of whether derivatives beyond the second order in the PDE can be included remains open.
Finally, the present approach could also be advantageous for other fields, besides quantum physics, where non-Gaussian high-dimensional Fokker-Planck equations are involved.

% ================= METHODS ======================
\section*{Methods}\label{sec:methods}

% ================= MODEL ================
\subsubsection*{Mappings quasi-Fokker-Planck equation: } The mappings between operators in the Lindbland master equation \cref{eq:LME} and quasi-FP equation for $p(\bm{x}$ \cref{eq:FP}
, see e.g. \cite{gardinerQuantumWorldUltraCold2014} are
$$s =\begin{cases}
			-1, \hspace{0.5cm}  p(\alpha,\alpha^*)=Q(\alpha,\alpha^*) & \text{Husimi-Q}\\
            0 \hspace{0.9cm}  p(\alpha,\alpha^*)=W(\alpha,\alpha^*)  & \text{Wigner}
		 \end{cases}$$
\begin{align}
    &a\rho &\Longleftrightarrow & &\left(\alpha - \frac{s-1}{2}\frac{\partial}{\partial \alpha^*}\right) p(\alpha,\alpha^*)\\
    &a^\dag\rho &\Longleftrightarrow & &\left(\alpha^*-\frac{s+1}{2}\frac{\partial}{\partial \alpha}\right)p(\alpha,\alpha^*)\\
    &\rho a  &\Longleftrightarrow & &\left(\alpha-\frac{s-1}{2}\frac{\partial}{\partial \alpha^*}\right)p(\alpha,\alpha^*)\\
    &\rho a^\dag &\Longleftrightarrow & &\left(\alpha^*-\frac{s-1}{2}\frac{\partial}{\partial \alpha}\right)p(\alpha,\alpha^*)
\end{align}
These can be applied iteratively to obtain the correct terms. As an example, for \cref{eq:CatHam} in combination with the dissipator $\mathcal{D}[\sqrt{\kappa}a]$, this yields the following quasi-FP equation for the Husimi-Q QPD, separating the real and imaginary part
\begin{align}
\begin{split}
    \label{eq:CatDecohQ_PDEmain}
   &  \dot{Q}(x_1, x_2) =\\
   &-\frac{\partial}{\partial x_1}\underbrace{ \left(\Delta x_2 - 2U x_2 (x_1^2 + x_2^2-2) -\frac{1}{2}\kappa x_1\right)}_{\nu_1}   Q(x_1, x_2)\\
    &-\frac{\partial}{\partial x_2}\underbrace{\left(-\Delta x_1 + 2U x_1 (x_1^2 + x_2^2-2)
-\frac{1}{2}\kappa x_2 \right)}_{\nu_2}  Q(x_1, x_2)\\
+&\bigg[\frac{\partial^2}{\partial x_1^2}\underbrace{\left(-U x_1 x_2 +\frac{1}{4}\kappa\right)}_{D_{11}}+ \frac{\partial^2}{\partial x_2^2}\underbrace{\left(U x_1 x_2 +\frac{1}{4}\kappa\right)}_{D_{22}} \\
      +&\frac{\partial^2}{\partial x_1 \partial x_2}\underbrace{\frac{1}{2}U(x_1^2-x_2^2)}_{D_{12}}+ \frac{\partial^2}{\partial x_2\partial x_1 }\underbrace{\frac{1}{2}U(x_1^2-x_2^2)}_{D_{21}} \bigg]Q(x_1, x_2),
\end{split}
\end{align}
with $\bm{x}=(\Re(\alpha), \Im(\alpha))$.

See SI for the remaining quasi-FP equations corresponding to the examples in the main text.

\subsubsection*{Details on the variational ansatz: }
We explore two ansätze to parameterize $ p(\bm{x} ,t)\rightarrow  p(\bm{x}; \bm{\theta}(t))$ , a complex Gaussian mixture (cGM) and a restricted Boltzmann machine (RBM) see \cref{fig:method}. The cGM consists of $R$ Gaussians $p(\bm{x};\bm{\theta}) = \sum_r^R c_r G_r(\mu_r,\Sigma_r;\bm{x})$, for $\bm{x} \in \mathbb{R}^{2M}$, where $c_r$ is the relative weight or normalization of each Gaussian and $G_r(\mu_r,\Sigma_r;\bm{x}) = \exp[-\frac{1}{2}(\bm{x}-\mu_r)^T\Sigma_r^{-1}(\bm{x}-\mu_r)]$, with complex mean $\mu_r$ and complex covariance matrix $\Sigma_r$.
Separating the real and imaginary part, the length of the parameter vector $\bm{\theta} = (\bm{\mu},\bm{\Sigma}, \bm{c})$, $N_\theta= R(4M+ 8M^2 + 1)$.
The Wigner and Q function of a cat state can be represented by the sum of four complex Gaussians (see the SI and Ref.~\cite{bourassaFastSimulationBosonic2021}). 

The RMB-based ansatz is defined as a tail-regularized 1-layer feed forward neural network: $p(\bm{x};\bm{\theta}) = K(\bm{x},\bm{\theta})F_\text{NN}(\bm{x};\bm{\theta}_\lambda)$, with $2M\beta$ hidden nodes with hyperbolic tangent activation function.
The coefficient $K(\bm{x},\bm{\lambda}) = \exp(- \sum_i \lambda_i^2 x_i^2)$ enforces the correct tail behaviour.
The parameters are $\bm{\theta} = (\bm{W},\bm{b},\bm{\lambda}$), where $\bm{W}$ are the network weights and $\bm{b}$ its biases, has length $N_\theta =\beta ((2M)^2 + 2M) + 2M$.
The initial condition for the evolution can be obtained by either fitting the ansatz of choice by means of gradient descent techniques to the discretized QPD corresponding to a given density matrix, or to an analytic expression. 
% See section \cref{sec:repPower} for the representative power of the ansatze.

\subsubsection*{Monte-Carlo sampling} 

Quantum-mechanical observables and quantities needed to compute the variational dynamics are computed using Monte-Carlo sampling of the QPD in the spirit of variational Monte Carlo.
If $p(\bm{x})\geq 0$ is a valid probability distribution, such as for the Husimi-Q QPD, we write quantum expectation values as 
\begin{equation}
    \label{eq:exp-val-def:Q}
    \expval*{\hat{O}} = \frac{\int Q(\bm{x}) O(\bm{x})\,d\bm{x}}{\int Q(\bm{x})\,d\bm{x}} = \mathbb{E}_{\bm{x}\sim Q(\bm{x})}\left[O(\bm{x})\right],
\end{equation}
where $O(\bm{x})$ is obtained from the mappings of operators to phase space (see Sec.A of the SI).
We explicitly write the denominator because the distribution $Q(\bm{x})$ is generally not normalized to 1.
The Wigner QPD, instead, can take on negative values and we cannot sample from it directly. 
Instead, we rewrite expectation values as statistical averages over samples drawn from $\abs{W(\bm{x})}$ as follows,

\begin{align}
\begin{split}
    \expval*{\hat{O}} &= \frac{\int W(\bm{x}) O(\bm{x})\,d\bm{x}}{\int W(\bm{x})\,d\bm{x}} \\
    &=\frac{\int \abs{W(\bm{x})} \,\text{sgn} \,(W(\bm{x})) O(\bm{x})\,d\bm{x}}{\int \abs{W(\bm{x})} \,\text{sgn} \,(W(\bm{x})) \,d\bm{x}} \\
    &= \frac{\mathbb{E}_{\bm{x}\sim \abs{W(\bm{x})}}\left[\text{sgn} \,(W(\bm{x}))  O(\bm{x})\right]  }{\mathbb{E}_{\bm{x}\sim \abs{W(\bm{x})}}\left[\text{sgn} \,(W(\bm{x}))  \right] }
    \label{eq:exp-val-def:W2}
\end{split}
\end{align}

These expectation values are then estimated by their sample mean over a polynomially-large set of samples obtained from a Markov-Chain Monte-Carlo sampling procedure (see Ref.~\cite{vicentiniNetKetMachineLearning2021} for details of the MCMC implementation employed).

New configurations for the chains are generated by the gaussian transition rule
\begin{equation}
    T(\bm{x}'|\bm{x}) = N(\bm{\mu}=\bm{x}, \sigma=10^{-1}),
\end{equation}
where $N(\bm{\mu}, \sigma)$ is a normal distribution with mean $\bm{\mu}$ and variance $\sigma$.  

%In a nutshell, tVMC methods  \cite{Carleo2017,rehTimeDependentVariationalPrinciple2021b,beccaQuantumMonteCarlo2017} sample the parameterized version $p(\bm{\theta};\bm{x})$, and evolve the parameters instead of the function. The distance $\mathbb{D}$ between the true evolution and the parameter evolution $\dot{\theta}$ up to first order is minimized ($\frac{ \partial \mathbb{D} }{\partial \dot{\theta}} =0 $) to find the optimal step $\dot{\theta}_\text{opt}$ 
%\begin{equation}%\label{eq:TDVP}
%   \dot{\theta}_\text{opt} = \argmin_{\dot{\theta}}  \infdiv*{p+\dot{p}dt }{p+ \sum_{k} \frac{\partial p}{\partial \theta}\dot{\theta}_{k}dt}
%\end{equation}
%Choosing a suitable distance such as the Kullback-Leiber, Hellinger or $L^2$ distance, this optimal step is given by the linear system of ODEs in \cref{eq:tdvp-equation-def} \cite{rehVariationalMonteCarlo2022a}.

The operators, sampling procedure and time step can be conveniently be implemented using the continuous variable back-end of NetKet \cite{vicentiniNetKetMachineLearning2021}, and utilizing its integrated parallelization options.

%\section*{Data availability}
%No external data was used in this study.
\section*{Code availability}
The code used in this study is available from the authors on reasonable request.

\bibliography{bib_PhaseSpacetVMC}% Produces the bibliography via BibTeX.

\section*{Acknowledgements}

The authors would like to thank Fabrizio Minganti for his valuable insights and critical reading of the manuscript, and thank David Schlegel for fruitful discussions.

\section*{Author Contributions}
D.E. and F.V. and V.S. designed the research; D.E. and F.V. developed the framework. D.E., F.V. and V.S. wrote the paper.
\section*{Competing Interests}
The authors declare no competing interests.

\clearpage
\pagenumbering{arabic}
\appendix
% ========== TERMS ========
\section{Operator mappings}\label{app:Terms}
We list the mappings between operators in the Lindblad master equation \cref{eq:LME} and quasi-FP equation \cref{eq:FP} for $p(\bm{x})$ 
, see e.g. \cite{gardinerQuantumWorldUltraCold2014} Ch. 16. 
$$
s =\begin{cases}
			-1, \hspace{0.5cm}  p(\alpha,\alpha^*)=Q(\alpha,\alpha^*) & \text{Husimi-Q}\\
            0 \hspace{0.9cm}  p(\alpha,\alpha^*)=W(\alpha,\alpha^*)  & \text{Wigner}\\
            +1,\hspace{0.5cm}   p(\alpha,\alpha^*)=P(\alpha,\alpha^*)  &  \text{Glauber P}
		 \end{cases}
$$
\begin{align*}
    &a\rho &\Longleftrightarrow & &\left(\alpha - \frac{s-1}{2}\frac{\partial}{\partial \alpha^*}\right) p(\alpha,\alpha^*)\\
    &a^\dag\rho &\Longleftrightarrow & &\left(\alpha^*-\frac{s+1}{2}\frac{\partial}{\partial \alpha}\right)p(\alpha,\alpha^*)\\
    &\rho a  &\Longleftrightarrow & &\left(\alpha-\frac{s-1}{2}\frac{\partial}{\partial \alpha^*}\right)p(\alpha,\alpha^*)\\
    &\rho a^\dag &\Longleftrightarrow & &\left(\alpha^*-\frac{s-1}{2}\frac{\partial}{\partial \alpha}\right)p(\alpha,\alpha^*)
\end{align*}
These can be applied iteratively to obtain the correct terms. Below we list the resulting complex evolution PDE for a number of common terms. To obtain  $p(\bm{x})$ in \cref{eq:FP}, one must separate the real and imaginary parts.
\subsection{Terms}
\subsubsection{Detuning}
\begin{itemize}
\item Hamiltonian term:
\begin{equation}
    H = \Delta a^\dag a 
\end{equation}
\item Wigner: 
\begin{equation}
    \dot{W}(\alpha, \alpha^*)  = i \Delta \left[ \frac{\partial}{\partial \alpha} \alpha
    -\frac{\partial}{\partial \alpha^*}\alpha^* \right] W(\alpha, \alpha^*) \\
\end{equation}
\item \text{Husimi-Q:} 
\begin{equation}
    \dot{Q}(\alpha, \alpha^*)  = i \Delta \left[ \frac{\partial}{\partial \alpha} \alpha
    -\frac{\partial}{\partial \alpha^*}\alpha^* \right] Q(\alpha, \alpha^*) \\
\end{equation}
\end{itemize}

% ==== NONLINEAR TERM  ==== 
\subsubsection{Kerr nonlinearity}
\begin{itemize}
\item Hamiltonian term .
\begin{equation}
    H = U (a^\dag a^\dag a a) 
\end{equation}

\item \text{Wigner:} \\
\begin{align}
\begin{split}
    &\dot{W}(\alpha, \alpha^*)  = \\
    &i U   \left[\frac{\partial}{\partial \alpha}(2 \alpha (|\alpha|^2-1))+\frac{\partial}{\partial \alpha^*}(-2\alpha^* (|\alpha|^2-1))\right] W(\alpha, \alpha^*)\\
    &+\underbrace{i U  \left[\frac{\partial^2}{\partial \alpha^*\alpha} \left(\frac{\partial}{\partial \alpha^{*}} \alpha^{*}- \frac{\partial}{\partial \alpha}\alpha \right)\right]}_\text{higher order} W(\alpha, \alpha^*) 
\end{split}
\end{align}

\item \text{Husimi-Q:} \\
\begin{align}
\begin{split}
    &\dot{Q}(\alpha, \alpha^*)  = \\
    &i U   \left[\frac{\partial}{\partial \alpha}(2 \alpha (|\alpha|^2-2))+\frac{\partial}{\partial \alpha^*}(-2\alpha^* (|\alpha|^2-2))\right] Q(\alpha, \alpha^*)\\
    +&i U   \left[\frac{\partial^2}{\partial \alpha^2} \alpha^2-\frac{\partial^2}{\partial \alpha^{*2}} \alpha^{*2}\right] Q(\alpha, \alpha^*) 
\end{split}
\end{align}

\end{itemize}

% ==== 1 photon Drive ==== 
\subsubsection{1 photon drive}
\begin{itemize}
\item Hamiltonian term (assuming $F$ is real).
\begin{equation}
    H = F (a^\dag +a) 
\end{equation}

\item \text{Wigner:} \\
\begin{equation}
    \dot{W}(\alpha, \alpha^*) =  iF \left[-\frac{\partial}{\partial \alpha}+\frac{\partial}{\partial \alpha^*}\right] W(\alpha, \alpha^*) \\
\end{equation}
\item \text{Husimi-Q:} \\
\begin{equation}
    \dot{Q}(\alpha, \alpha^*)  =  iF \left[-\frac{\partial}{\partial \alpha}+\frac{\partial}{\partial \alpha^*}\right] Q(\alpha, \alpha^*) \\
\end{equation}
\end{itemize}
% ==== 2 photon Drive ==== 
\subsubsection{2 photon drive}
\begin{itemize}
\item Hamiltonian term (assuming $G$ is real).
\begin{equation}
    H = G (a a + a^\dag a^\dag) 
\end{equation}

\item \text{Wigner:} \\
\begin{equation}
    \dot{W}(\alpha, \alpha^*) =  iG \left[-\frac{\partial}{\partial \alpha}2\alpha^*-\frac{\partial}{\partial \alpha^*}2\alpha\right] W(\alpha, \alpha^*) \\
\end{equation}

\item \text{Husimi-Q:} \\
\begin{equation}
    \dot{Q}(\alpha, \alpha^*)  =  iG \left[-\frac{\partial}{\partial \alpha}2\alpha^*-\frac{\partial}{\partial \alpha^*}2\alpha+\frac{\partial^2}{\partial \alpha^2}-\frac{\partial^2}{\partial \alpha^{*2}}\right] Q(\alpha, \alpha^*) \\
\end{equation}

\end{itemize}

% ==== 1 photon Dissipation ==== 
\subsubsection{1 photon dissipation}
\begin{itemize}
\item Jump operators:
\begin{align}
    C_1 &= \sqrt{\gamma(1+n_\text{th})}a = \sqrt{\kappa_a} a \\
    C_2 &= \sqrt{\gamma n_\text{th}}a^\dag = \sqrt{\kappa_b} a^\dag 
\end{align}
\item Lindbladian term $C_1$:
\begin{equation}
    \dot{\rho}(t) =  D[\sqrt{\kappa_a} a ]
\end{equation}
with jump operator
\begin{equation}\label{eq:dissipOperatorJ2}
   D[\sqrt{\kappa_a} a ] =\frac{\kappa_a}{2} \left[2a \rho(t) a ^\dag - \rho(t)a^{\dag}a-a^{\dag}a\rho(t)\right] 
\end{equation}
\item \text{Wigner:} \\
\begin{align}
\begin{split}
    &\dot{W}(\alpha, \alpha^*) = \\
    &\kappa_a \left[\frac{1}{2}\frac{\partial}{\partial \alpha}\alpha+\frac{1}{2}\frac{\partial}{\partial \alpha^*}\alpha^*+\frac{1}{4}\frac{\partial^2}{\partial \alpha \partial \alpha^*}+\frac{1}{4}\frac{\partial^2}{\partial \alpha^*\partial \alpha}\right]  W(\alpha, \alpha^*) 
\end{split}
\end{align}

\item \text{Husimi-Q:} \\
\begin{align}
\begin{split}
&\dot{Q}(\alpha, \alpha^*) = \\
&\kappa_a \left[\frac{1}{2}\frac{\partial}{\partial \alpha}\alpha+\frac{1}{2}\frac{\partial}{\partial \alpha^*}\alpha^*+\frac{1}{2}\frac{\partial^2}{\partial \alpha \partial \alpha^*}+\frac{1}{2}\frac{\partial^2}{\partial \alpha^*\partial \alpha}\right]  Q (\alpha, \alpha^*) 
\end{split}
\end{align}

\item Lindbladian term $C_2$:
\begin{equation}
    \dot{\rho}(t) =  D[\sqrt{\kappa_b} a^\dag ]
\end{equation}
with jump operator
\begin{equation}
   D[\sqrt{\kappa_b} a^\dag  ] =\frac{\kappa_b}{2} \left[2a^\dag \rho(t) a - \rho(t)aa^\dag -aa^\dag \rho(t)\right] 
\end{equation}
\item \text{Wigner:} \\
\begin{align}
\begin{split}
    &\dot{W}(\alpha, \alpha^*) =\\
    &\kappa_b \left[-\frac{1}{2}\frac{\partial}{\partial \alpha}\alpha-\frac{1}{2}\frac{\partial}{\partial \alpha^*}\alpha^*+\frac{1}{4}\frac{\partial^2}{\partial \alpha \partial \alpha^*}+\frac{1}{4}\frac{\partial^2}{\partial \alpha^*\partial \alpha}\right]  W(\alpha, \alpha^*)  \end{split}
\end{align}
\item \text{Husimi-Q:} \\
\begin{align}
\begin{split}
&\dot{Q}(\alpha, \alpha^*) = \\
&\kappa_b \left[-\frac{1}{2}\frac{\partial}{\partial \alpha}\alpha-\frac{1}{2}\frac{\partial}{\partial \alpha^*}\alpha^*+\frac{1}{2}\frac{\partial^2}{\partial \alpha \partial \alpha^*}+\frac{1}{2}\frac{\partial^2}{\partial \alpha^*\partial \alpha}\right]  Q(\alpha, \alpha^*) 
\end{split}
\end{align}

\end{itemize}

% ========= 2 photon Dissipation ==========

\subsubsection{2 photon dissipation}
\begin{itemize}
\item Jump operators:
\begin{equation}
    C=  \sqrt{\eta} a a
   \end{equation}
\item Lindbladian term $C$:
\begin{equation}
    \dot{\rho}(t) =  D[\sqrt{\eta} a a ]
\end{equation}
with jump operator
\begin{equation}\label{eq:dissipOperatorJ}
   D[\sqrt{\eta}  a ] =\frac{\eta}{2} \left[2a a \rho(t) a ^\dag a ^\dag - \rho(t)a^{\dag} a ^\dag a a-a^{\dag}a ^\dag a a\rho(t)\right] 
\end{equation}
\item \text{Wigner:} \\
\begin{align}
\begin{split}
    \dot{W}(\alpha, \alpha^*) &= \eta \Bigg[ \Bigg. \frac{\partial}{\partial \alpha}(|\alpha|^2-\alpha)+\frac{\partial}{\partial \alpha^*}(|\alpha|^2-\alpha^*)\\
    &+\frac{1}{2}\frac{\partial^2}{\partial \alpha \partial \alpha^*}(|\alpha|^2-1) + 
    \frac{1}{2}\frac{\partial^2}{\partial \alpha^*\partial \alpha }(|\alpha|^2-1)\\
    &+\underbrace{\frac{\partial^3}{\partial \alpha^*\partial \alpha^2}\alpha+\frac{\partial^3}{\partial \alpha^{*2}\partial \alpha} \alpha^*}_\text{higher order}\Bigg. \Bigg]  W (\alpha, \alpha^*) 
\end{split}
\end{align}
\item \text{Husimi-Q:} \\
\begin{align}
\begin{split}\label{eq:2Photdissip}
    \dot{Q}(\alpha, \alpha^*) &= \eta \Bigg[ \Bigg. \frac{\partial}{\partial \alpha}(\alpha|\alpha|^2-2\alpha)+\frac{\partial}{\partial \alpha^*}(\alpha^*|\alpha|^2-2\alpha^*)\\
    &+2\frac{\partial^2}{\partial \alpha \partial \alpha^*}|\alpha|^2 + 
    2\frac{\partial^2}{\partial \alpha^*\partial \alpha }|\alpha|^2\\
        &+
    \frac{\partial^2}{\partial \alpha^2 }\alpha^2 + \frac{\partial^2}{\partial \alpha^{*2} }\alpha^{*2}\\
    &+\underbrace{\frac{\partial^3}{\partial \alpha^*\partial \alpha^2}\alpha+\frac{\partial^3}{\partial \alpha^{*2}\partial \alpha} \alpha^*}_\text{higher order}\Bigg. \Bigg]  Q (\alpha, \alpha^*) 
\end{split}
\end{align}
\end{itemize}
% ==== Hopping n ==== 
\subsubsection{Nearest neighbor hopping}
\begin{itemize}
\item Hamiltonian term:
\begin{equation}
    H = J \sum_{j=1}^{N-1}(a_{j+1}^\dag a_j   + a_j^\dag a_{j+1} ) 
\end{equation}
with $N$ the number of modes
\item Lindbladian term:
\begin{equation}
    \dot{\rho}(t) = -i[H,\rho(t)] 
\end{equation}

\item \text{Wigner evolution :} 
\begin{equation}
    \dot{W}(\bm{\alpha}, \bm{\alpha}^*)  = \sum_{j=1}^{N-1} i J_j \left[ \frac{\partial}{\partial \alpha_j} \alpha_{j+1}
    -\frac{\partial}{\partial \alpha_j^*}\alpha_{j+1}^* \right] W(\bm{\alpha}, \bm{\alpha}^*) \\
\end{equation}

\item \text{Husimi-Q evolution :} 
\begin{equation}
    \dot{Q}(\bm{\alpha}, \bm{\alpha}^*) = \sum_{j=1}^{N-1} i J_j \left[ \frac{\partial}{\partial \alpha_j} \alpha_{j+1}
    -\frac{\partial}{\partial \alpha_j^*}\alpha_{j+1}^* \right] Q(\bm{\alpha}, \bm{\alpha}^*) \\
\end{equation}

\end{itemize}

\clearpage

% =================== APPENDIX ANALYTIC EXPRESSIONS =============================
\section{Analytic expressions of Wigner and Q functions:}\label{app:exprCatCoh}
\begin{itemize}
    \item  A coherent state $\ket{\alpha}$ in one mode can be expressed by as a Gaussian or normal distribution in both the Husimi-Q and Wigner representation:
\begin{align}
    \begin{split}
            p(\bm{x})_\text{coh} &=  \mathcal{G} (\bm{\mu},\bm{\Sigma};\bm{x}) \\
            &= \frac{1}{\sqrt{(2\pi)^k |\boldsymbol\Sigma|}}\exp\left(-\frac 1 2 ({\mathbf x}-{\boldsymbol\mu})^\mathrm{T}{\boldsymbol\Sigma}^{-1}({\mathbf x}-{\boldsymbol\mu})\right)\\
            &= w \exp\left(-\frac 1 2 ({\mathbf x}-{\boldsymbol\mu})^\mathrm{T}{\boldsymbol\Sigma}^{-1}({\mathbf x}-{\boldsymbol\mu})\right) 
    \end{split}
\end{align}

with  $\bm{\mu}=[\Re(\alpha), \Im(\alpha) ]$ and  covariance matrix $\bm{\Sigma}=\frac{1-s}{4}\bm{I}$, where $s=-1$ for $Q(\bm{x})$, and $s=0$ for  $W(\bm{x})$.
\vspace{.5 cm}

\item A cat state $\ket{\mathcal{C}_\pm} = \ket{\alpha} \pm \ket{-\alpha}$ can be described exactly by a summation of four complex Gaussians, two for each lobe and two for the interference fringes, as derived by \cite{bourassaFastSimulationBosonic2021} for the Wigner function. 

\begin{equation}\label{eq:catW}
    p(\bm{x})_\text{cat}=  \mathcal{G}_-(\bm{x}) + \mathcal{G}_+(\bm{x})  + \mathcal{G}_z(\bm{x})  + \mathcal{G}_{\bar{z}}(\bm{x}) 
\end{equation}

We derive the same holds for the Husimi-Q function, be it with the same factor $2$ difference in the covariance matrix, as the coherent state.
See \cref{tab:cat_coeff} for the coefficients of the Gaussian mixture.

Due to the smaller variance, the interference fringes in between to coherent states are much smaller in the Q representation than the Wigner representation.

\item 
For multi mode initial conditions, the tensor product can be taken for the coefficients.
\end{itemize}

%\begin{table}[]
%    \centering
%    \begin{tabular}{c|c|c|c|c}
%     $G$    &    $w_i$                                  & $\bm{\mu}_i$                &            $W: \bm{\Sigma}_i$   &            $Q: \bm{\Sigma}_i$\\
%     \hline
%    $G_+$   &  $\mathcal{N} $                                    & [$\Re(\alpha)$,$\Im(\alpha)$] &      $\frac{1}{4}\bm{I}$&       $\frac{1}{2}\bm{I}$ \\
%    $G_-$   &   $\mathcal{N}  $                                             & [$\Re(-\alpha)$,$\Im(-\alpha)$] &   $\frac{1}{4}\bm{I}$&       $\frac{1}{2}\bm{I}$ \\
%    $G_z$   &  $ \mathcal{N} e^{i k \pi}e^{-2|\alpha|^2}    $          & [$i\Im(\alpha)$,$-i\Re(\alpha)$] &         $\frac{1}{4}\bm{I}$&       $\frac{1}{2}\bm{I}$ \\
%    $G_{\Bar{z}}$   & $  \mathcal{N} e^{i k \pi}e^{-2|\alpha|^2} $         & [$-i\Im(\alpha)$,$+i\Re(\alpha)$] &        $\frac{1}{4}\bm{I}$&       $\frac{1}{2}\bm{I}$ \\
%    \end{tabular}
%    \caption{Coefficients for the 4-component Gaussian mixture that exactly represents a cat state, where the defintion of the Wigner and the Q function only differens in the value of the covariance matrix. $\mathcal{N}$ is a normalization constant that ensures $ \sum_i^R w_i =1$, with $R=4$. Parameter $k$ determines if the cat state is an even ($k=1$) or odd ($k=0$).}
 %   \label{tab:cat_coeff}
%\end{table}

\begin{table}[]
    \centering
    \begin{tabular}{c|c|c|c|c}
     $G$    &    $w_i$                                  & $\bm{\mu}_i$                &           $\bm{\Sigma}_i$       \\      
     \hline
    $G_+$   &  $\mathcal{N} $                                    & [$\Re(\alpha)$,$\Im(\alpha)$] &      $\frac{1-s}{4}\bm{I}$ \\
    $G_-$   &   $\mathcal{N}  $                                             & [$\Re(-\alpha)$,$\Im(-\alpha)$] &   $\frac{1-s}{4}\bm{I}$ \\
    $G_z$   &  $ \mathcal{N} e^{i k \pi}e^{-2|\alpha|^2}    $          & [$i\Im(\alpha)$,$-i\Re(\alpha)$] &         $\frac{1-s}{4}\bm{I}$\\
    $G_{\Bar{z}}$   & $  \mathcal{N} e^{i k \pi}e^{-2|\alpha|^2} $         & [$-i\Im(\alpha)$,$+i\Re(\alpha)$] &        $\frac{1-s}{4}\bm{I}$ \\
    \end{tabular}
    \caption{Coefficients for the 4-component Gaussian mixture that exactly represents a cat state, where the definition of the Wigner and the Q function only differs in the value of the covariance matrix. $\mathcal{N}$ is a normalization constant that ensures $ \sum_i^R w_i =1$, with $R=4$. Parameter $k$ determines if the cat state is even ($k=1$) or odd ($k=0$).}
    \label{tab:cat_coeff}
\end{table}

%\section{TDVP S-matrix and F-vector}
%For more details, see %\cite{rehTimeDependentVariationalPrinciple2021b}
%Defining $\mathcal{O}_k(x) = \frac{\partial \log p %(x)}{\partial \theta_k}$ 
%\begin{align}
%    F_k &= \int p(x) \frac{\partial \log p(x)}{\partial t}\frac{\partial \log p(x)}{\partial \theta_k}dx\\
%    &= \langle \mathcal{O}_k(x)  \frac{\partial \log p(x)}{\partial t} \rangle
%\end{align}
%and
%\begin{align}
%    S_{kj} &=  \int p(x) \frac{\partial \log p(x)}{\partial \theta_k}\frac{\partial \log p(x)}{\partial \theta_j}dx \\
 %   &=\langle \mathcal{O}_k(x)  \mathcal{O}_j(x) \rangle
%\end{align}

%where $\frac{\partial \log p(x)}{\partial t}=\frac{1}{p(x)}\frac{\partial p(x)}{\partial t}$ is obtained from the operators in the quasi FP equation \cref{eq:FP}. 

\section{Examples manuscript}
\begin{itemize}
    \item \textit{1 cavity cat decoherence}

      Hamiltonian and dissipator:
        \begin{equation}\label{eq:CatHamApp}
    H = \Delta a^\dag a  - U (a^\dag a^\dag a a)  , \quad  D[\sqrt{\kappa}a]
\end{equation}
Fokker-Planck like PDE:
\begin{align}
\begin{split}\label{eq:CatDecoh_PDE}
    \dot{p}(\alpha, \alpha^*) =& i \Delta \left[ \frac{\partial}{\partial \alpha} \alpha
    -\frac{\partial}{\partial \alpha^*}\alpha^* \right] p(\alpha, \alpha^*)\\
       +&i U   \bigg[\frac{\partial}{\partial \alpha}(2 \alpha (|\alpha|^2-(s-1)))\\
       +&\frac{\partial}{\partial \alpha^*}(-2\alpha^* (|\alpha|^2-(s-1)))\bigg]  p(\alpha, \alpha^*)\\
       +&i U s\left[\frac{\partial^2}{\partial^2 \alpha}(\alpha^{*2})+\frac{\partial^2}{\partial^2 \alpha^*}\alpha^{2}\right]  p(\alpha, \alpha^*)\\
        +&\underbrace{i U (1-s^2) \left[\frac{\partial^2}{\partial \alpha^*\alpha} \left(\frac{\partial}{\partial \alpha^{*}} \alpha^{*}- \frac{\partial}{\partial \alpha}\alpha \right)\right]}_\text{higher order} p(\alpha, \alpha^*) \\
      +& \kappa \bigg[\frac{1}{2}\left(\frac{\partial}{\partial \alpha}\alpha+\frac{\partial}{\partial \alpha^*}\alpha^*\right)\\
      +&\frac{1-s}{4}\left(\frac{\partial^2}{\partial \alpha \partial \alpha^*}+\frac{\partial^2}{\partial \alpha^*\partial \alpha}\right)\bigg]  p(\alpha, \alpha^*) .
\end{split}
\end{align}
  
%    \begin{widetext}

%\end{widetext}
\item \textit{1 cavity bistable initial condition}

      Hamiltonian and dissipator:
        \begin{equation}\label{eq:BistabApp}
    H = \Delta a^\dag a  - U (a^\dag a^\dag a a) + F(a^\dag +a)  , \quad  D[\sqrt{\kappa}a]
\end{equation}
Fokker-Planck like PDE:
\begin{align}
\begin{split}
    \dot{Q}(\alpha, \alpha^*)  =& i \Delta \left[ \frac{\partial}{\partial \alpha} \alpha
    -\frac{\partial}{\partial \alpha^*}\alpha^* \right]  Q(\alpha, \alpha^*)\\
    +&i U   \bigg[\frac{\partial}{\partial \alpha}(2 \alpha (|\alpha|^2-2))\\
    +&\frac{\partial}{\partial \alpha^*}(-2\alpha^* (|\alpha|^2-2))\bigg] Q(\alpha, \alpha^*)\\
    +&i U   \left[\frac{\partial^2}{\partial \alpha^2} \alpha^2-\frac{\partial^2}{\partial \alpha^{*2}} \alpha^{*2}\right] Q(\alpha, \alpha^*) \\
   +& iF \left[-\frac{\partial}{\partial \alpha}+\frac{\partial}{\partial \alpha^*}\right] Q(\alpha, \alpha^*) \\
      +&\kappa_a \bigg[\frac{1}{2}\frac{\partial}{\partial \alpha}\alpha
      +\frac{1}{2}\frac{\partial}{\partial \alpha^*}\alpha^*\\
      +&\frac{1}{2}\frac{\partial^2}{\partial \alpha \partial \alpha^*}
      \frac{1}{2}\frac{\partial^2}{\partial \alpha^*\partial \alpha}\bigg]  Q(\alpha, \alpha^*)
\end{split}
\end{align}

\item \textit{1 cavity competition nonlinarity and dissipation}

      Hamiltonian and dissipator:
        \begin{equation}\label{eq:BM}
    H =  U (a^\dag a^\dag a a), \quad  D[\sqrt{\kappa}a]
\end{equation}
Fokker-Planck like PDE:
\begin{align}
\begin{split}
    \dot{Q}(\alpha, \alpha^*)  =&i U   \bigg[\frac{\partial}{\partial \alpha}(2 \alpha (|\alpha|^2-2))\\
    +&\frac{\partial}{\partial \alpha^*}(-2\alpha^* (|\alpha|^2-2))\bigg] Q(\alpha, \alpha^*)\\
    +&i U   \left[\frac{\partial^2}{\partial \alpha^2} \alpha^2-\frac{\partial^2}{\partial \alpha^{*2}} \alpha^{*2}\right] Q(\alpha, \alpha^*) \\
      +&\kappa_a \bigg[\frac{1}{2}\frac{\partial}{\partial \alpha}\alpha+\frac{1}{2}\frac{\partial}{\partial \alpha^*}\alpha^*\\
      +&\frac{1}{2}\frac{\partial^2}{\partial \alpha \partial \alpha^*}+\frac{1}{2}\frac{\partial^2}{\partial \alpha^*\partial \alpha}\bigg]  Q(\alpha, \alpha^*) 
\end{split}
\end{align}

\item \textit{3 cavity hopping and forcing} ($M=3$) 
 Hamiltonian and dissipators:
\begin{align}
\begin{split}\label{eq:ham:3modeHopApp}
     H = &\sum_i^M (\Delta_i a_i^\dag a_i   - U (a_i^\dag a_i^\dag a_i a_i)) +  F (a_1^\dag+ a_1)  \\ \\
    &+ J\left[ \sum_{i=1}^{M-1}(a_{i+1}^\dag a_i   + a_i^\dag a_{i+1} )   +  a_{1}^\dag a_M   + a_M^\dag a_{1}  \right] \\
    , & \quad \sum_i^M  D[\sqrt{\kappa}a_i]
\end{split}
\end{align}

Fokker-Planck like PDE:
\begin{align}
\begin{split}
    \dot{Q}(\bm{\alpha} , \bm{\alpha} ^*) = &\sum_{j=1}^{M=3} \bigg( i \Delta \left[ \frac{\partial}{\partial \alpha_j} \alpha_j-\frac{\partial}{\partial \alpha_j^*}\alpha_j^* \right]\\
   +&i U   \bigg[\frac{\partial}{\partial \alpha_j}(2 \alpha_j (|\alpha_j|^2-2))\\
   +&\frac{\partial}{\partial \alpha_j^*}(-2\alpha_j^* (|\alpha_j|^2-2))\bigg] \\
    +&i U   \left[\frac{\partial^2}{\partial \alpha_j^2} \alpha_j^2-\frac{\partial^2}{\partial \alpha_j^{*2}} \alpha_j^{*2}\right] \\
      +&\kappa_a \bigg[\frac{1}{2}\frac{\partial}{\partial \alpha_j}\alpha_j+\frac{1}{2}\frac{\partial}{\partial \alpha_j^*}\alpha_j^*\\
      +&\frac{1}{2}\frac{\partial^2}{\partial \alpha_j \partial \alpha_j^*}+\frac{1}{2}\frac{\partial^2}{\partial \alpha_j^*\partial \alpha_j}\bigg] \bigg) Q(\bm{\alpha} , \bm{\alpha} ^*)  \\
      +&iF \left[-\frac{\partial}{\partial \alpha_1}+\frac{\partial}{\partial \alpha_1^*}\right]Q(\bm{\alpha} , \bm{\alpha} ^*)\\
      +&iJ\Bigg(\sum_{j=1}^{M-1}   \left[ \frac{\partial}{\partial \alpha_j} \alpha_{j+1}
    -\frac{\partial}{\partial \alpha_j^*}\alpha_{j+1}^* \right] \\
    +& \left[ \frac{\partial}{\partial \alpha_M} \alpha_{1}
    -\frac{\partial}{\partial \alpha_M^*}\alpha_{1}^* \right] \Bigg) Q(\bm{\alpha} , \bm{\alpha} ^*)
    \end{split}
    \end{align}

\item \textit{6 cavity hopping} ($M=6$)\newline
Hamiltonian and dissipators:
\begin{align}
\begin{split}\label{eq:ham:6modeHopApp}
     H = &\sum_i^M (\Delta_i a_i^\dag a_i   - U (a_i^\dag a_i^\dag a_i a_i))    \\
    &+ J\left[ \sum_{i=1}^{M-1}(a_{i+1}^\dag a_i   + a_i^\dag a_{i+1} )   +  a_{1}^\dag a_M   + a_M^\dag a_{1}  \right] \\
    , & \quad \sum_i^M  D[\sqrt{\kappa}a_i]
\end{split}
\end{align}

Fokker-Planck like PDE:
\begin{align}
\begin{split}
    \dot{Q}&(\bm{\alpha} , \bm{\alpha} ^*) = \sum_{j=1}^{M=6} \Bigg(\\
    & i \Delta \left[ \frac{\partial}{\partial \alpha_j} \alpha_j
    -\frac{\partial}{\partial \alpha_j^*}\alpha_j^* \right] \\
    +&i U   \bigg[\frac{\partial}{\partial \alpha_j}(2 \alpha_j (|\alpha_j|^2-2))\\
    +&\frac{\partial}{\partial \alpha_j^*}(-2\alpha_j^* (|\alpha_j|^2-2))\bigg] \\
    +&i U   \left[\frac{\partial^2}{\partial \alpha_j^2} \alpha_j^2-\frac{\partial^2}{\partial \alpha_j^{*2}} \alpha_j^{*2}\right] \\
      +&\kappa_a \bigg[\frac{1}{2}\frac{\partial}{\partial \alpha_j}\alpha_j+\frac{1}{2}\frac{\partial}{\partial \alpha_j^*}\alpha_j^*\\
      +&\frac{1}{2}\frac{\partial^2}{\partial \alpha_j \partial \alpha_j^*}+\frac{1}{2}\frac{\partial^2}{\partial \alpha_j^*\partial \alpha_j}\bigg] \Bigg) Q(\bm{\alpha} , \bm{\alpha} ^*) \\
      +&iJ\Bigg(\sum_{j=1}^{M-1}   \left[ \frac{\partial}{\partial \alpha_j} \alpha_{j+1}
    -\frac{\partial}{\partial \alpha_j^*}\alpha_{j+1}^* \right] \\
    +& \left[ \frac{\partial}{\partial \alpha_M} \alpha_{1}
    -\frac{\partial}{\partial \alpha_M^*}\alpha_{1}^* \right] \Bigg) Q(\bm{\alpha} , \bm{\alpha} ^*)
    \end{split}
    \end{align}
\end{itemize}
While we kept the parameters $U,F,J,\Delta$ and $\kappa_a$ the same for each cavity for simplicity, they can be set individually without an extra numerical effort.

\clearpage
\section{Phase space dynamics and effect of sample size}
\label{appendix:graph}
\begin{figure*}[ht]
\includegraphics[width=18cm]{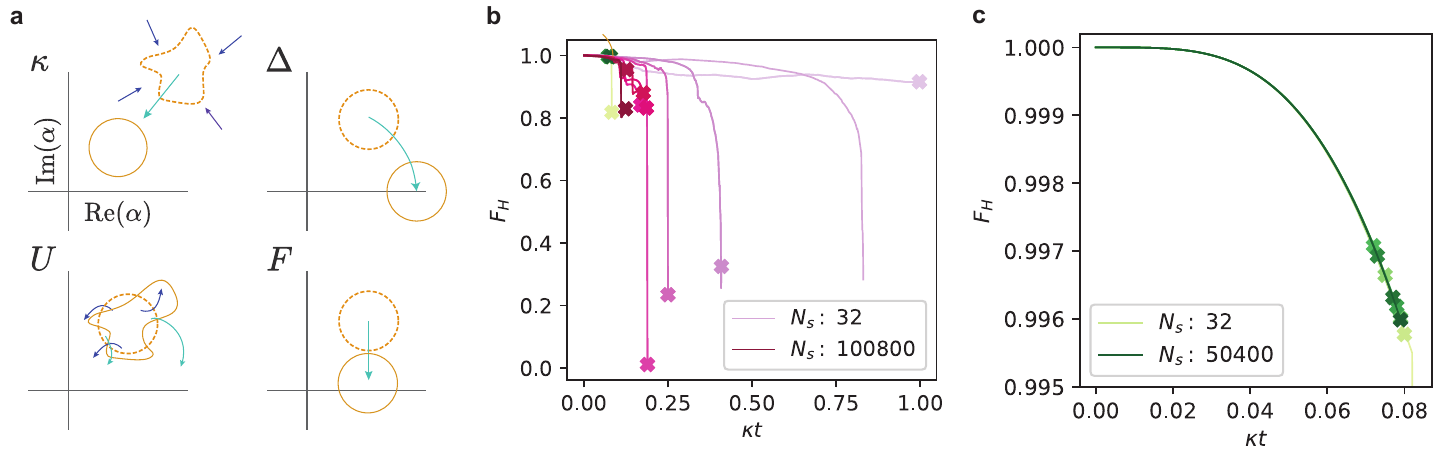}
\centering
\caption{ \textbf{a}Sketch of the individual effect of various terms in \cref{eq:FP} on the evolution. Dashed line indicates the initial shape, solid line the final shape after some time $t$. Purple arrows indicate diffusion, green arrows drift.  \textbf{b} $F_\text{H}$  as a function of time for  for RMB (purple) for different sample sizes ($N_s = [    16   ,  32    , 64  ,  128    ,256   ,1008 ,  2016,  10080 , 20160, 100800]$ (light to dark), and cGM ansatz (green)  $N_s = [  16 ,   32   , 64  , 128  , 256  , 512 , 1008,  5040 ,10080 ,50400]$ (light to dark). The final point before the solver breaks down in indicated with a cross.  \textbf{c} Zoom-in for $F_H$ for cGM ansatz as a function of time. Note the different vertical scale. \label{fig:MethodLimits}   }  
\end{figure*}

In \cref{fig:MethodLimits}(a) we evaluate the effects of various terms in \cref{eq:HamBistable} and $D[\sqrt{\kappa}a]$ on the phase space, by examining their behavior in \cref{eq:FP}
\begin{enumerate}
    \item The dissipation term ($\kappa$)  contains a drift component that drives towards the origin (green arrows). The diffusion matrix is diagonal, and its magnitude is related to the variance of a coherent state in the Wigner and Q representation. That is, if the field is a coherent state (a Gaussian with the correct variance), the diffusion (purple arrows) has no effect. However, if there is an initially odd-shaped field (dashed lines), this term will drive it to a Gaussian shape.  Together, the drift and diffusion terms drive towards a Gaussian state in the origin, i.e. the vacuum. 
    \item  The detuning term  ($\Delta$) contains only a drift term with opposite sign for $\alpha$ and $\alpha^*$, thus inciting a circular motion around the origin.
    \item The single photon drive term  ($F$) is determined by the complex value of $F$. With the other terms present, when real and positive, this induces a drift towards $\alpha^*$. 

    \item Unlike the previous terms, the nonlinear term ($U$) consists of a drift operator that contains nonlinear coordinates in its argument, indicated by the nonlinear/curved arrows in \cref{fig:MethodLimits}a, which can cause deformation to the shape of the function. That is, when starting out with a Gaussian field (dashed line) that only has a first and second moment (mean and variance), this term introduces higher moments, or non-Gaussianity. This is unlike the first three terms, which are either Gaussian preserving, or actively driving towards a Gaussian such as the dissipation term. 
    Specifically for a coherent state away from the origin, the distance to the origin determines the drift velocity (in contrast to for instance the detuning term), causing a 'swirling' deformation displayed in \cref{fig:catdecoh}b,f).
    
    Another term that contain nonlinearity in the coordinates is two photon dissipation, as shown in SI 1.
\end{enumerate}

\Cref{fig:MethodLimits}(a,b) examines the influence of the number of sample $N_\text{s}$  to estimate the $S$ matrix and $F$ vector in \cref{eq:tdvp-forces-definition,eq:tdvp-qgt-definition}. The breakdown point, where the parameterized solution starts to deviate too much from the true solution and the solver breaks down, is marked with a cross for different $N_s$. The number of samples are indicated by a color range of light (low $N_s$) to dark (high $N_s$).

For both ansätze the breakdown point shows no clear dependence on $N_\text{s}$, after a sufficient number of samples.
The restrictive cGM ansatz (zoom-in in panel c) can be accurately sampled with low $N_s$. Notice the different fidelity range on the axes.

\end{document}